\documentclass{appolb}

\usepackage{pstricks}
\usepackage[dvips,final]{graphicx}
\usepackage{amsmath,amssymb,bm,pifont}

\usepackage{graphicx}

\begin{document}
\title{Worldline techniques and QCD observables%
\thanks{Presented at Light-Cone 2012, July 08-13, 2012, Cracow, Poland. \\
        Dedicated to the memory of Walter Gl\"ockle, teacher, friend,
        and colleague.}%
}
\author{N. G. Stefanis
\address{Institut f\"{u}r Theoretische Physik II,
         Ruhr-Universit\"{a}t Bochum,
         D-44780 Bochum, Germany
         }
\\
}
\maketitle
\begin{abstract}
This report attempts to capture the essential workings of gauge links
(Wilson lines) inside gauge-invariant formulations of parton
distribution functions in QCD and gain some deeper insight into their
key (renormalization) properties.
We show, in particular, that the one-loop anomalous dimension of the
Cherednikov-Stefanis quark TMD PDF is in the lightcone gauge $A^+=0$,
combined with the Mandelstam--Leibbrandt pole prescription, the same
as that obtained in the special covariant gauge $a=-3$, leaving no
uncanceled rapidity singularities.
\end{abstract}

\section{Introduction}
\label{sec:intro}
The standard way to remove the gauge-dependence of nonlocal correlators
in gauge theories, like QCD, is to include path-ordered exponential
factors of the gauge field that are absent in the original Lagrangian
of a local quantum field theory.
These operators are in general contour-dependent and give rise to new
divergences, called in modern jargon rapidity divergences, that are not
related to the standard singularities --- ultraviolet (UV) and infrared
(IR) --- that appear in Feynman diagrams.
They originate from contour irregularities caused by topological
obstructions --- endpoints, cusps, self-crossings --- of the gauge
contours entering the exponents of the gauge links and affect the
renormalization properties of the QCD correlators \cite{Pol79}.
The trouble is, such divergences have to be regularized and there is
great ambiguity in adopting an appropriate subtraction procedure
within a valid factorization scheme.
Moreover, the use of the lightcone gauge quantization depends
crucially on the adopted boundary conditions imposed on the gluon
propagator in order to treat the gauge links at infinity.

These issues will be considered here in some detail.
In Sec.\ \ref{eq:mesonic-correlator}, we will address the one-loop
virtual corrections of the quark propagator in a general covariant
gauge within two different gauge-invariant schemes:
The Mandelstam formalism \cite{Man68YM} and the $z$-field algorithm
\cite{HJS77,GN79}.
Then, in Sec.\ \ref{sec:tmd-correlators}, we will discuss --- as an
example of a transverse-momentum dependent (TMD) parton distribution
function (PDF) --- the quark in a quark TMD PDF, $f_{q/q}$, using the
lightcone gauge subject to various boundary conditions on the gluon
propagator.
Finally, the conclusions will be presented in Sec.\ \ref{sec:concl}.

\section{Mesonic correlator}
\label{sec:meson-correlator}
In this section, we consider a gauge-invariant mesonic type
correlator in QCD, {\it viz.},
\begin{equation}
    M(x_1, x_2|\Gamma)
=
  \bar{q}_{i}(x_2) [x_{2}, x_{1}|\Gamma]_{ij} q_{j}(x_1) \, ,
\label{eq:mesonic-correlator}
\end{equation}
where
\begin{equation}
[x_{2}, x_{1}|\Gamma]_{ij}
=
  \mathcal{P}\exp \left[
                        ig \int_{x_1[\Gamma]}^{x_2} dz_{\mu}A_{a}^{\mu}(z)t_{ij}^{a}
                  \right]
\label{eq:gauge-link}
\end{equation}
is a path-ordered gauge link along some arbitrary contour $\Gamma$
between $x_1$ and $x_2$ and
$t_{ij}^{a} = (1/2)\lambda_{ij}^{a}$
are the Gell-Mann matrices of $SU(3)_{c}$.
In principle, any contour between the points $x_1$ and $x_2$ is
admissible.
Contour obstructions give rise to rapidity divergences and hence
contribute to the anomalous dimension of the correlator.
For our considerations in this section, we assume that the contour is
smooth.
Because only the two endpoints entail rapidity divergences, it is
sufficient to employ the straight line $\bar\Gamma$ joining $x_1$
and $x_2$.
The reason is that other features of smooth contours, e.g., their
length, do not affect the renormalization properties of the
twist-two mesonic correlator and hence $\bar\Gamma$ can be used as a
subtraction contour for its renormalization for all members of the
universality class of smooth contours.
Differences in the definition of the mesonic correlator for different
smooth contours show up at the next higher twist level
\cite{CKS10}.\footnote{I thank Sergey Mikhailov for discussions on
this point.}

The gauge link can be treated in two different ways.
One can evaluate the path-ordered exponential as a power series in
the coupling using the Mandelstam formalism \cite{Man68YM}.
This approach \cite{Ste83} will be on focus in the first subsection
below.
Another option is to apply the $z$-field algorithm \cite{HJS77,GN79},
which is based on an effective Lagrangian describing the interaction
of one-dimensional auxiliary Fermion fields with the gluon field and
trade contours for ``particle trajectories''.
Fully quantized results are finally obtained by performing a functional
integral over the $z$-field fluctuations to get the mesonic correlator
in second quantization.
Our presentation here follows the analysis of \cite{CD80} with more
details to be given in a future publication.

\subsection{Mandelstam formalism}
\label{subsec:mandelstam}
\begin{figure}[t!]
\centerline{
\includegraphics[width=0.28\textwidth]{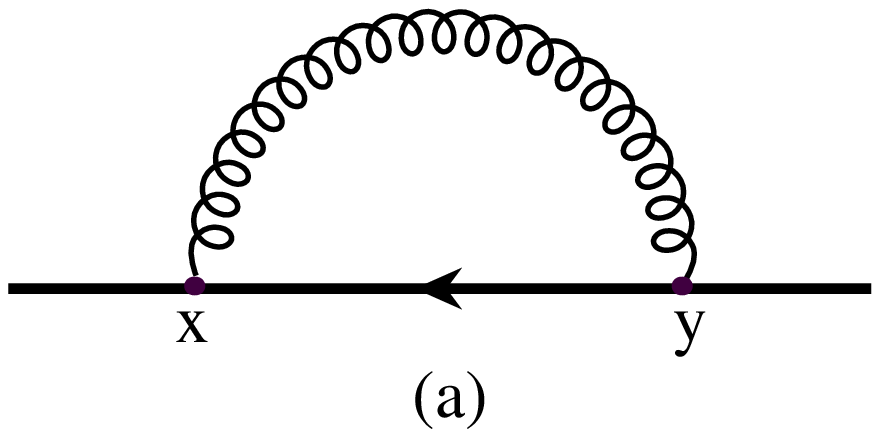}~~~~~%
\includegraphics[width=0.28\textwidth]{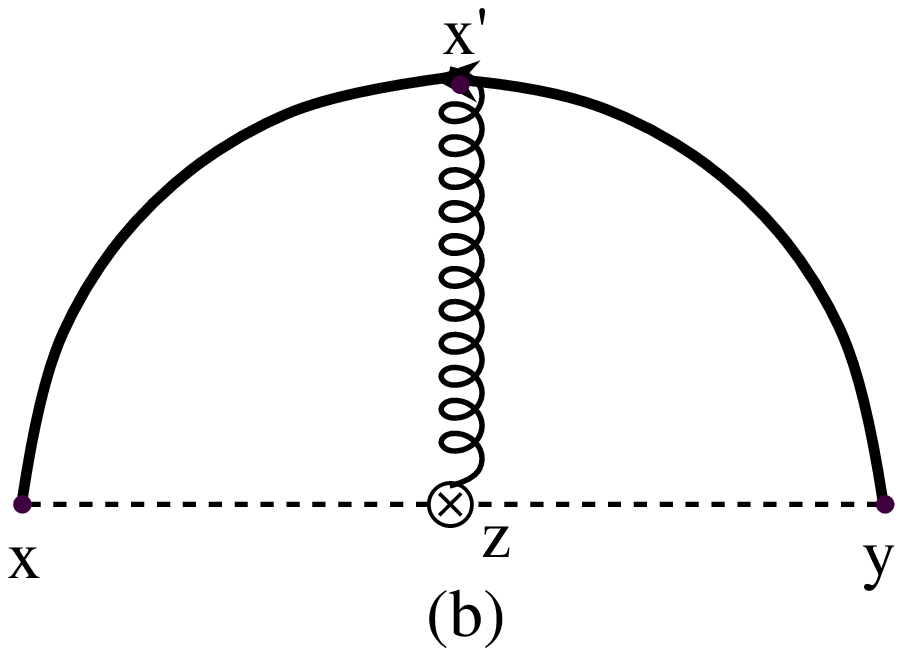}~~~~~%
\includegraphics[width=0.28\textwidth]{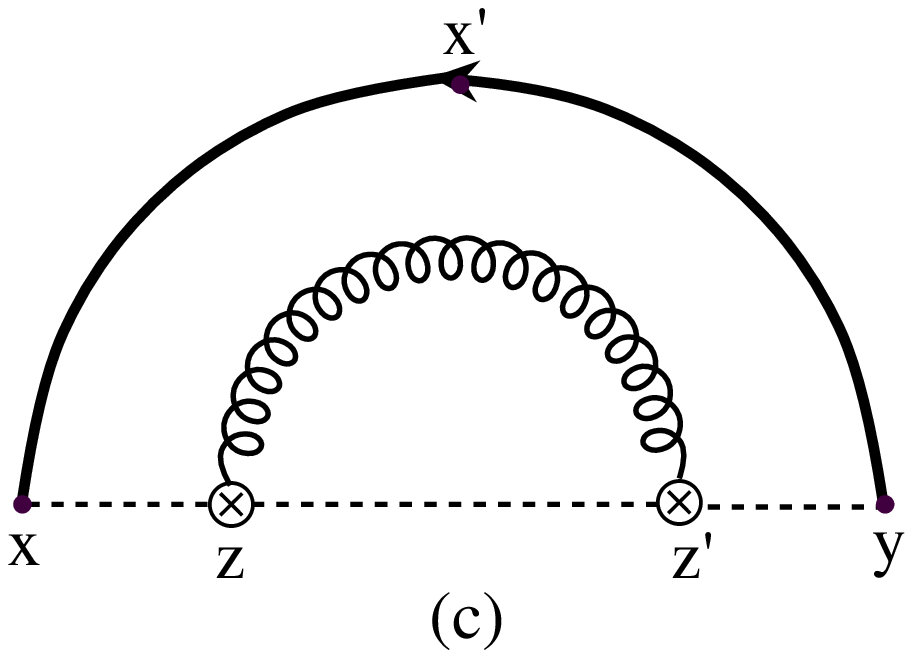}}
  \caption{\label{fig:self-energy-Man}
    One-loop contributions to the vacuum expectation value of the
    mesonic correlator in Eq.\ (\ref{eq:GF-Man}) within the Mandelstam
    formalism.
    (a) Usual quark self-energy with gluon attachment denoted by a
    curly line;
    (b) and (c) contributions due to
    Eqs.\ (\ref{eq:cross-talk}) and (\ref{eq:link-self-energy}),
    respectively.
    Curved solid lines denote quark propagators, whereas the gauge
    contours are represented by dashed straight lines with the symbol
    $\otimes$ denoting line integrals along them.}
\end{figure}

We carry out the vacuum expectation value of $M(x_1, x_2|\Gamma)$ to the
second order of the unrenormalized coupling constant $g$, i.e.,
\begin{eqnarray}
  \langle 0 | T \left(
                \bar{q}(x_2) [x_{2}, x_{1}|\Gamma] q(x_1)
                \right)
  \! | 0 \rangle^{(2)}
& \!\! = \!\! &
  \langle 0 | T \left(
                \bar{q}(x_2) [x_{2}, x_{1}|\Gamma] q(x_1)
                \right)
  \! | 0 \rangle^{(2)}
\nonumber \\
&& \!\! +
  \langle 0 | T \left(
                \bar{q}(x_2) [x_{2}, x_{1}|\Gamma]^{(1)} q(x_1)
                \right)
  \! | 0 \rangle^{(1)}
\nonumber \\
&& \!\! +
  \langle 0 | T \left(
                \bar{q}(x_2) [x_{2}, x_{1}|\Gamma]^{(2)} q(x_1)
                \right)
  \! | 0 \rangle^{(0)}
\label{eq:GF-Man}
\end{eqnarray}
with each contribution evaluated to the appropriate order.
The first term is the usual gauge-dependent quark self-energy to
$\mathcal{O}(g^2)$, whereas the second and the third term represent
the $\mathcal{O}(g^1)$ and the $\mathcal{O}(g^2)$ contributions
stemming from the gauge link (termed in \cite{Ste83} the connector),
respectively:
\begin{eqnarray}
  [x_{2}, x_{1}|\Gamma]^{(1)}
& = &
  i g \int_{x_{1}[\Gamma]}^{x_2} dz_{\mu} A^{\mu}(z) \ ,
\label{eq:cross-talk}
\end{eqnarray}
\begin{eqnarray}
  [x_{2}, x_{1}|\Gamma]^{(2)}
& = &
  \frac{(i g)^2}{2!} \int_{x_{1}[\Gamma]}^{x_2} dz_{\mu}
  \int_{x_{1}[\Gamma]}^{x_2} dz_{\nu}^{\prime}
  \mathcal{P}\left(A^{\mu}(z)A^{\nu}(z^{\prime})\right) \ .
\label{eq:link-self-energy}
\end{eqnarray}
The three terms displayed in Eq.\ (\ref{eq:GF-Man}) are shown
graphically in Fig.\ \ref{fig:self-energy-Man}.

It was shown in \cite{Ste83} that the rapidity divergences entailed by
the endpoints in diagrams (b) and (c) in Fig.\
\ref{fig:self-energy-Man} can be controlled by dimensional
regularization with no need to involve additional regulators.
The explicit calculation can be found there.
Here we only quote the final results recalling that we are dealing with
quark fields that are Heisenberg field operators in the interaction
picture (where they are free operators) so that one has to perform all
Wick contractions in
\begin{eqnarray}
  \mathcal{M}(x_1, x_2|\Gamma)
=
  \left\langle
         0\left| T \!\left(
         \bar{q}_{0}(x_2)
         [x_2, x_1 |\Gamma]
         q_{0}(x_1)
         \exp \left( \! i \! \int d^{4}y \mathcal{L}_{\rm int}(y)\right) \!
             \right)
  \right| 0 \right\rangle .
\label{eq:Heisenberg-field}
\end{eqnarray}
The mesonic correlator in a general covariant gauge $a$ can be written
in the form
\begin{eqnarray}
  \mathcal{M}^{(i,j)}(p|\bar{\Gamma})
=
  \mathcal{M}_{1}^{(a,b)}(p) + (1 - a) \mathcal{M}_{2}^{(a,b)}(p)
\label{eq:generic-corr}
\end{eqnarray}
with
\begin{eqnarray}
  D_{\rm F}^{\mu\nu}(k)
=
  -\frac{g^{\mu\nu}}{k^2 + i\epsilon}
  +(1-a) \frac{k^\mu k^\nu/k^2}{k^2 + i\epsilon} \ , \quad
S_{\rm F}(p)
=
  \frac{1}{\hat{p}-m+i\epsilon} \ ,
\label{eq:propagators}
\end{eqnarray}
where the first superscript in $\mathcal{M}^{(i,j)}$ indicates the
order of the expansion of the gauge link, whereas the second one
denotes the order of the coupling taken into account in the Wick
contractions.
One finds \cite{Ste83} that the connector parts are interrelated:
$
 \mathcal{M}_{2}^{(1,1)}(p)
=
 - 2 \mathcal{M}_{2}^{(2,0)}(p),
$
while the remaining part of
$
 \mathcal{M}_{2}^{(1,1)}(p)$
cancels the $(1-a)$ contribution of
$
 \mathcal{M}_{2}^{(0,2)}(p),
$
so that
$\mathcal{M}^{(2)}(p|\bar{\Gamma})$
is gauge-parameter independent and multiplicatively renormalizable:
\begin{eqnarray}
  \mathcal{M}(p|\bar{\Gamma})|_{\rm ren}
=
  Z_{\rm hybrid}^{-1} \mathcal{M}(p|\bar{\Gamma})|_{\rm bare} \ .
\label{eq:ren-corr}
\end{eqnarray}

Using dimensional regularization we obtain in the $\overline{\rm MS}$
scheme the following renormalization constants ($D=4-\epsilon$)
\cite{Ste83}
\begin{eqnarray}
  Z_{\rm hybrid}
& = &
  Z_{2q} Z_{\rm con}
=
  1 + \frac{3g^2}{8\pi^2} C_{\rm F} \frac{1}{\epsilon} + \mathcal{O}(g^4)
\\
  Z_{\rm con}
& = &
  1 + \frac{g^2}{8\pi^2} C_{\rm F}(3+a) \frac{1}{\epsilon} + \mathcal{O}(g^4) \ , \quad
\\
  Z_{2q}
& = &
  1 - \frac{g^2}{8\pi^2} C_{\rm F} a \frac{1}{\epsilon} + \mathcal{O}(g^4) \ ,
\label{eq:ren-constants-man}
\end{eqnarray}
where
$C_{\rm F}=(N_{c}^{2}-1)/(2N_{\rm c})=4/3$ for $N_{\rm c}=3$
The associated anomalous dimensions
($
  \gamma
=
  \frac{1}{2} \mu\frac{\partial}{\partial\mu} \ln Z
$)
are given by
\begin{eqnarray}
  \gamma_{\rm hybrid}
& = &
  \gamma_{2q} + \gamma_{\rm con}
=
  - \frac{3 g^2}{16\pi^2}C_{\rm F} + \mathcal{O}(g^4)
\\
  \gamma_{\rm con}
& = &
  - \frac{g^2}{16\pi^2}C_{\rm F}(3 + a) + \mathcal{O}(g^4)\ , \quad
\\
  \gamma_{2q}
& = &
  \frac{g^2}{16\pi^2}C_{\rm F} a + \mathcal{O}(g^4)
\label{eq:anom-dim-man}
\end{eqnarray}
and bear no dependence on the geometric features of the contour, e.g.,
its length or its derivatives $d^nx(\tau)/(d\tau)^n$.

\subsection{$z$-field formalism}
\label{subsec:z-fields}
Though the correlator $\mathcal{M}(x_1, x_2|\Gamma)$ is multiplicatively
renormalizable, it cannot be written \emph{contour-independently} in the
factorized form
$
  M(x_1, x_2|\Gamma)
=
  \bar{Q}(x_2|\Gamma_2)Q(x_1|\Gamma_1)
$
starting from the QCD Lagrangian in terms of the Mandelstam field
$
 Q(x_1| \Gamma_1)
=
  \mathcal{P}\exp\left[
                       i g \int_{-\infty[{\Gamma_1}]}^{x_1} dz_\mu A_{a}^{\mu}(z) t_a
                 \right]
  q(x_1).
$
However, for smooth contours one can factorize the connector
$[x_1, x_2|\Gamma]$ according to the algebraic identity
$[B,A|\Gamma]=[B,C|\Gamma_2][C,A|\Gamma_1]$ with
$\Gamma_1\cup \Gamma_2=\Gamma$ and then shift $C\to \infty$.
This trick allows factorization and multiplicative renormalization,
i.e.,
$
  Q(x|\Gamma)|_{\rm ren}
=
  Z_{\rm hybrid}^{-1/2} Q(x|\Gamma)|_{\rm bare}.
$
\begin{figure}[t!]
\centerline{
\includegraphics[width=0.7\textwidth]{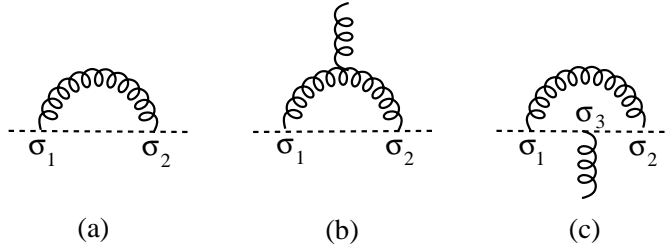}}%
\caption{\label{fig:z-field-corrections}
    One-loop contributions to the mesonic correlator in the $z$-field
    formalism: (a) $z$-field (broken line) self-energy; (b) and (c)
    vertex 
    corrections from
    $\bar{z}A^{\mu}\frac{dx_{\mu}(\sigma)}{d\sigma}z$.}
\end{figure}
A more far-sighted option is to trade gauge contours in favor of
``trajectories'' of fictitious particles described by the following
effective Lagrangian
\begin{eqnarray}
  \mathcal{L}_{\rm eff}
=
  \int_{0}^{1} d\sigma  \left[
                                \bar{z}(\sigma)\frac{\partial}{\partial\sigma}z(\sigma)
                              + ig \bar{z}(\sigma)\frac{dx_{\mu}(\sigma)}{d\sigma}
                                A^\mu(x(\sigma))z(\sigma)
                              + i\bar{\lambda}z
                              + i\bar{z}\lambda
                          \right]
\label{eq:eff-Lag}
\end{eqnarray}
and supplement $\mathcal{L}_{\rm QCD}$ by two additional terms
pertaining to two extra Feynman rules: one for the $z$-field propagator
and the other for the $z$-field--gluon vertex (see \cite{CD80}).
This allows one to write the gauge link as a path integral:
\begin{eqnarray}
  \mathcal{P} \exp \left[
                         i g \int_{\sigma_1}^{\sigma_2} d\sigma
                         A_{a}^{\mu}(x(\sigma)) t^{a} \frac{dx_{\mu}}{d\sigma}
                   \right]_{ij}
\equiv
  \langle 0|z_{i}(\sigma_2) \bar{z}_{j}(\sigma_1)|0 \rangle \ ,
\label{eq:z-field-ident}
\end{eqnarray}
where
$
 \langle \mathcal{Q}\rangle
=
  \int \mathcal{D}\bar{z} \mathcal{D}z \mathcal{Q}
  \exp \left[
       \int_{\sigma_1}^{\sigma_2} d\sigma \bar{z}(\sigma) D_\sigma z(\sigma)
       \right].
$
What is more, performing now the calculation of the radiative
corrections to the mesonic correlator within this approach
(see Fig.\ \ref{fig:z-field-corrections}), one finds
at $\mathcal{O}(g^2)$ that the \emph{local} combination
$\bar{z}(\sigma_x)q(x)$
(analogous to the \emph{nonlocal} field $Q(x|\Gamma)$ in the Mandelstam
formalism)
gets renormalized by the renormalization constant \cite{CD80}
$
 Z_{\rm CF}^{-1/2}
=
 \tilde{Z}_{\rm CF}^{-1/2} Z_{2q}^{-1/2} Z_{3z}^{1/2},
$
where
$
 Z_{3z}
=
 1 + \frac{g^2}{8\pi^2} C_{\rm F} (3-a) \frac{1}{\epsilon}
$
resembles $Z_{\rm con}$, yielding to the same anomalous dimensions
as before.
As a result, the Slavnov-Taylor identities to $\mathcal{O}(g^2)$ are
fulfilled ($C_{\rm A}=N_{\rm c}=3$):
\begin{equation}
 \frac{Z_{\rm con}Z_{1q}}{Z_{\rm hybrid}}
=
  \frac{Z_{3}}{Z_{1}}
=
  \frac{\tilde{Z}_{3}}{\tilde{Z}_{1}}
=
  \frac{Z_{3z}}{Z_{1z}}
=
  1 + \frac{C_{\rm A}g^2}{32\pi^2}(3+a)\frac{1}{\epsilon} \ .
\label{eq:ST-id-man}
\end{equation}
One can now use the $z$-field formalism and perform a short-distance
expansion of $M(x_1, x_2|\Gamma)$ for $x_1\longrightarrow x_2$:
\begin{eqnarray}
  M(x_1, x_2|\Gamma)
& = &
  \bar{q}(x_2) [x_{2}, x_{1}|\Gamma] q(x_1)
=
  \bar{q}(x_2) z(\sigma_2) \bar{z}(\sigma_1) q(x_1)
\\ \nonumber
& \approx &
  \sum_{N,i} C_{N}^{(i)}(z^2,g,C)
  z^{\mu_1} \ldots z^{\mu_N}
  \mathcal{O}_{z^{\mu_1} \ldots z^{\mu_N}}^{(i)}(x,C) \ ,
\label{eq:expansion}
\end{eqnarray}
where
$(x_2 - x_1)^2=z^2 \simeq 0$, $z^\mu \neq 0$
and the composite non-singlet quark operator of lowest twist reads
\begin{eqnarray}
  \mathcal{O}_{z^{\mu_1} \ldots z^{\mu_N}}^{(i)}(x,C)
=
  \bar{q}(x) \gamma_{\mu_1}
  \overleftrightarrow{D}_{\mu_1} \ldots \overleftrightarrow{D}_{\mu_N}
  [x,x|C]
  q(x) \ .
\label{eq:comp-op}
\end{eqnarray}
Here $[x,x|C]$ is a gauge link along the closed loop
$C=\Gamma \cup \Gamma'$
and the short-distance expansion is valid for any point $x$ because
the smooth contour $\Gamma'$ can be \emph{stretched} to $\infty$ by
virtue of the independence of the renormalization constants on
$L(\Gamma')$.
An immediate important conclusion is that in the special gauge
$a=-3$ all contour- (or $z$-field-) related divergences cancel
among themselves so that the residual renormalization effects
can be absorbed into $Z_3$, while the Ward identity $Z_{1q}=Z_{2q}$
is preserved like in QED.
It is expected that going to the next higher loop, one will obtain
a similar result but for $a^*=-3 + \mathcal{O}(g^2)$ as found in
\cite{Mik99} in the context of multiloop contributions to the
nonsinglet QCD evolution equations.

\section{Gauge-invariant correlators for TMD PDFs}
\label{sec:tmd-correlators}
With such issues in mind, let us focus attention on the question of
the appropriate definition of a TMD PDF (e.g., \cite{CR12} and
references cited therein).
In \cite{CS07} we have shown that $f_{q/q}(x,\bm{k}_{\perp})$ given
in \cite{BJY02} cannot be regularized completely using dimensional
regularization in the lightcone gauge $A^{+}=(A \cdot n^{-})=0$ with
$(n^-)^2=0$ in conjunction with the retarded (ret), advanced (adv),
or principal-value (PV) pole prescription on the gluon propagator:
\begin{equation}
   D_{\mu\nu}^{\rm LC} (q)
=
   \frac{-i}{q^2 - \lambda^2 + i0} \left( g_{\mu\nu}
  -\frac{q_\mu n^-_\nu + q_\nu n^-_\mu}{[q^+]}\right) \ ,
\label{eq:gluon-prop}
\end{equation}
where
\begin{eqnarray}
  \frac{1}{[q^+]}\Bigg|_{\rm Ret/Adv}
=
  \frac{1}{q^+ \pm i \eta} \ \ \ , \ \ \
  \frac{1}{[q^+]}\Bigg|_{\rm PV}
=
  \frac{1}{2} \left[ \frac{1}{q^+ + i \eta} + \frac{1}{q^+ - i \eta} \right] \ .
\label{eq:adv-ret-pv}
\end{eqnarray}
The reason is that the residue of the $\epsilon$ pole contains a
rapidity divergence that entails an extra anomalous dimension
and calls for an additional subtraction procedure.
This can be achieved in terms of a soft factor $R$ that has to be
included into the definition of the TMD PDF \cite{CS07}.
Its anomalous dimension $\gamma_R$ serves to cancel $\delta\gamma$
with the effect that the anomalous dimension of $f_{q/q}^{R}$
coincides with the result one would obtain in a covariant gauge
for a direct smooth contour between the two field points
i.e., Eq.\ (13).
It turns out \cite{CS07} that this anomalous-dimension artifact has
at one loop the same structure as the universal anomalous dimension
of a cusped contour \cite{KR87},
$
 \delta\gamma
=
 -\frac{\alpha_s}{\pi} C_{\rm F} \ln \frac{\eta}{p^+},
$
which becomes infinite when $\eta \to 0$.
For such $q^-$-independent pole prescriptions the gluon propagator
is not transverse:
$n_\mu  D^{\mu\nu} \neq 0$.
On the other hand, it was shown in \cite{CS09} that using instead the
$q^-$-dependent Mandelstam--Leibbrandt (ML) \cite{ML-pre} pole
prescription
\begin{eqnarray}
  D_{\rm ML}^{\mu\nu}(q^2)
=
  \frac{i}{q^2}
  \left[
        -g^{\mu\nu}
        + \frac{q^\mu n^\nu + q^\nu n^\mu}{[q^+]}
  \right]; ~~~
  \frac{1}{[q^+]_{\rm ML}}
= \left\{
  \begin{array}{ll}
  & \!\!\!\!\!\!\! \frac{1}{q^+ + i 0 q^-}     \\
  & \!\!\!\!\!\!\! \frac{q^-}{q^+ q^- + i0}
  \
  \end{array} \right.
\label{eq:ML-pre}
\end{eqnarray}
one has
$n_\mu  D_{\rm ML}^{\mu\nu} = 0$.
The opposite claims by Collins in \cite{Collins:2011ca} are likely
the result of misunderstanding, as his equation (15) shows the gluon
propagator subject to the Principal Value pole prescription, i.e.,
Eq.\ (\ref{eq:adv-ret-pv}).
This gluon propagator is indeed not transverse.
The same applies to those obtained with the advanced or retarded pole
prescription, because in all these cases the transverse gauge field
$
  \mathbf{A}^\perp (\infty^-; \bm{\xi}_\perp)
= \frac{g}{4\pi} C_{\infty}
   \bm{\nabla}^\perp \ln \Lambda |\bm{\xi}_\perp|
$
is not purely transverse (in contrast to the ML case) but depends
through
$C_\infty = \{ 0 (\rm adv), -1 (ret), -\frac{1}{2} (PV) \}$
on the imposed boundary conditions.

Another issue raised by Collins in \cite{Collins:2011ca} is whether
the definition of $f_{q/q}^{R}$ contains uncanceled divergences
originating from the self-energy of the gauge links.
We have shown in \cite{CS07,CS09} that all UV divergences from the
$q$ momentum integrations can be regularized dimensionally and give
$1/\epsilon$ poles, while collinear poles are controlled by the
quark virtuality $p^2<0$, and IR singularities are regularized by
an auxiliary gluon mass $\lambda$ that drops out at the end.
On the other hand, overlapping divergences have, in general, to be
cured by a subtraction procedure encoded in the soft renormalization
factor $R$ and appear in $\ln \frac{\eta}{p^+}$ in terms of an
auxiliary mass $\eta$.
In the $A^+=0$ gauge all diagrams with gluon attachments to the
gauge links (longitudinal or transverse) either vanish identically
or cancel partly against contributions from cross-talk diagrams with
gluon attachments between the quark line and a gauge link.
For the adv, ret, and PV pole prescriptions, all remaining divergences
are taken care of by the soft factor rendering $f_{q/a}^{R}$ regular.
Employing the ML prescription, one gets a result that is reminiscent of
that obtained in the special covariant gauge $a=-3$ in which all
rapidity divergences emerging from the endpoints of smooth contours
cancel among themselves.
At higher loops, the gauge $a=-3$ will receive $\alpha_s$ corrections
with coefficients that can be determined by demanding the validity of
the QED-like Ward identity
$Z_{1q}=Z_{2q}$.

\section{Conclusions}
\label{sec:concl}
We have shown the multiplicative renormalization of the gauge-invariant
mesonic vacuum correlator in the \emph{nonlocal} Mandelstam approach
for smooth contours and found that to $\mathcal{O}(g^2)$ it is equivalent
to the result obtained in the \emph{local} $z$-field effective formalism.
At one loop, all contour- (or $z$-field) related rapidity divergences
cancel among themselves using the special gauge $a=-3$.
We argued that a proper definition of the quark TMD PDF must account for
the appropriate subtraction of a rapidity divergence that overlaps with
the usual UV singularities and cannot be regularized dimensionally.
Employing the lightcone gauge with $q^-$-independent pole prescriptions,
this can be achieved via a soft renormalization factor \cite{CS07}.
The imposition of the $q^-$-dependent Mandelstam--Leibbrandt
prescription removes all rapidity divergences and reproduces at one loop 
the results of the special covariant gauge $a=-3$
with the soft factor reducing to unity.



\begin{thebibliography}{10}

\bibitem{Pol79}
A.M.\ Polyakov,
{\it Nucl.\ Phys.} {\bf B164}, 171 (1979).

\bibitem{Man68YM}
S.\ Mandelstam,
{\it Phys. Rev.} {\bf 175}, 1580 (1968).

\bibitem{HJS77}
M.B.\ Halpern, A.\ Jevicki, P.\ Senjanovic,
{\it Phys. Rev.} {\bf D16}, 2476 (1977).

\bibitem{GN79}
J.-L.\ Gervais, A.\ Neveu,
{\it Nucl.\ Phys.} {\bf B163}, 189 (1980).

\bibitem{CKS10}
I.O.\ Cherednikov, A.I.\ Karanikas, N.G.\ Stefanis
{\it Nucl.\ Phys.} {\bf B840}, 379 (2010).

\bibitem{Ste83}
N.G.\ Stefanis,
{\it Nuovo Cim.} {\bf A83}, 205 (1984).

\bibitem{CD80}
N.S. Craigie, H. Dorn,
{\it Nucl.\ Phys.} {\bf B185}, 204 (1981).

\bibitem{Mik99}
S.V.\ Mikhailov,
{\it Phys.\ Rev.} {\bf D62}, 034002 (2000);
{\it Phys.\ Lett.} {\bf 431}, 387 (1998).

\bibitem{CR12}
J.C.\ Collins, T.C.\ Rogers,
arXiv:1210.2100 [hep-ph].

\bibitem{CS07}
I.O.\ Cherednikov, N.G.\ Stefanis,
{\it Phys.\ Rev.} {\bf D77}, 094001 (2008);
{\it Nucl.\ Phys.} {\bf B802}, 146 (2008);
N.G.\ Stefanis, I.O.\ Cherednikov,
{\it Mod.\ Phys.\ Lett.} {\bf A24}, 2913 (2009).

\bibitem{BJY02}
A.V.\ Belitsky, X. Ji, F. Yuan,
{\it Nucl.\ Phys.} {\bf B656}, 165 (2003).

\bibitem{KR87}
G.P.\ Korchemsky, A.V.\ Radyushkin,
{\it Nucl.\ Phys.} {\bf B283}, 342 (1987).

\bibitem{CS09}
I.O.\ Cherednikov, N.G.\ Stefanis,
{\it Phys.\ Rev.} {\bf D80}, 054008 (2009).

\bibitem{ML-pre}
S.~Mandelstam,
{\it Nucl.\ Phys.} {\bf B213}, 149 (1983);
G.~Leibbrandt, S.L.~Nyeo,
{\it Phys.\ Lett.} {\bf B140}, 417 (1984).

\bibitem{Collins:2011ca}
J.C.\ Collins,
{\it Int.\ J.\ Mod.\ Phys.\ Conf.\ Ser.} {\bf 4}, 85 (2011).

\end{thebibliography}

\end{document}